\documentclass[11pt,twoside]{article}
\usepackage{newpasp}
\usepackage{graphicx}
\def\Teff{T_{\rm ef\kern-.05em f}} 
\def\vinf{v_\infty}
\def\Mdot{\dot M}
\def\Msun{M_\odot}
\def\Rsun{R_\odot}
\def\Lsun{L_\odot}

\begin{document}
\title{Radiation driven atmospheres of O-type stars:\\
constraints on the mass--luminosity relation of\\
central stars of planetary nebulae}
\author{A.\,W.\,A.\,Pauldrach(1), T.\,L.\,Hoffmann(1), R.\,H.\,M\'endez(1,2)}
\affil{(1)~Institut f\"ur Astronomie und Astrophysik der
Universit\"at M\"unchen,\\
Scheinerstra{\ss}e~1, 81679~M\"unchen, Germany;\\
(2)~Institute for Astronomy, University of Hawaii,\\
2680~Woodlawn Drive, Honolulu, HI~96822, U.S.A.}

\begin{abstract}
Recent advances in the modelling of stellar winds driven by
radiation pressure make it possible to fit many wind-sensitive
features in the UV spectra of hot stars, opening the way for a
hydrodynamically consistent determination of stellar radii,
masses, and luminosities from the UV spectrum alone.
It is thus no longer necessary to assume a theoretical
mass--luminosity relation.
As the method has been shown to work for massive O~stars,
we are now able to test predictions from the post-AGB
evolutionary calculations quantitatively for the first time.
Here we present the first rather surprising consequences of
using the new generation of model atmospheres for the
analysis of a sample of central stars of planetary nebulae.
\end{abstract}

\section{Introduction}

A lot of work on model atmospheres of PN central stars
(CSPNs in what follows) has been motivated by the desire
to obtain information about the basic properties of CSPNs
(surface temperature, mass, luminosity, abundances), so as to be
able to test predictions from post-AGB evolutionary calculations.
The earlier efforts, based on plane-parallel non-LTE models,
could not achieve a completely independent test, in the following sense:
since the plane-parallel model fits to H and He photospheric
absorption lines can only produce information about surface
temperature, He abundance and $\log g$, we cannot derive
stellar masses or luminosities, but only $L/M$ ratios.
This is exactly the same problem we face when dealing with
low-gravity early-type ``supergiant'' stars at high Galactic latitudes:
are they luminous and massive, or are they evolving away from the AGB?
We need some independent evidence to settle the issue -- for example,
the distance to the star.
Unfortunately, we lack reliable distances to most CSPNs.

So what could be done was to plot the positions of CSPNs in the
$\log g$--$\log \Teff$ diagram, and compare them with plots of
post-AGB tracks, translated from the $\log L$--$\log \Teff$ diagram.
After doing this translation it is possible to read the
stellar mass in the $\log g$--$\log \Teff$ diagram.
From this, we can derive $L$ and, if we know the visual
dereddened apparent magnitude, a so-called ``spectroscopic distance''.
All this work, however, is based on {\em assuming\/} that
the evolutionary models give us the correct relation
between stellar mass and luminosity.
It is not a real test of the evolutionary models,
but only a consistency check.

In the last 10 years there has been a lot of progress in the
modelling of stellar winds driven by radiation pressure.
Many CSPNs show spectroscopic evidence of winds, in the form
of emission lines in the visible spectrum and especially
P-Cygni-type profiles of resonance lines in the ultraviolet
between 1000 and 2000\,\AA.

The existence of these wind features provides both
a challenge and an opportunity.
The challenge is to model them.
The opportunity is to use the information about the
geometrical extension of the atmosphere and the forces of
gravity and radiative pressure, implicit in the wind profiles
(from which the terminal velocity and mass loss rate
can be derived), to obtain the physical size of the star,
which is the key to derive the stellar luminosity and mass.
(The idea is described together with a first application by
Pauldrach et al.~1988.)
Thus the successful modelling of the wind features opens the way
for a real test of the mass--luminosity relation of CSPNs.

In this review we would like to present the current situation
of the project and the rather surprising results obtained up to now.
Section 2 describes a model analysis of a massive Population I star,
using state-of-the-art hydrodynamically consistent, spherically
symmetric model atmospheres to demonstrate the power of the technique
and to show how successfully we can reproduce the ionizing fluxes
and observed spectra of such stars.
In Section 3 we introduce the wind-momentum--luminosity relation
and describe a previous attempt to determine whether or not
CSPNs follow this relation.
In Section 4 we show two examples of the application of our
hydrodynamically consistent wind models to CSPNs
(fits to IUE and HST spectra and derived parameters).
In Section 5 we add the results from 6 other CSPNs
similarly analyzed and discuss the consequences.

\section{Modelling a massive O supergiant: $\alpha$~Cam}

The analysis method is described in detail by Pauldrach et
al.~(2001).\footnote{
This paper is also available on the Web, at\\
http://www.usm.uni-muenchen.de/people/adi/adi.html.}
Here we can only give a brief summary.
The analysis method is based on modelling a homogeneous, stationary,
extended, outflowing, spherically symmetric radiation driven atmosphere.
A complete model atmosphere calculation involves solving the
hydrodynamics and the NLTE problem (rate equations, radiative transfer).
The solution of the total interdependent system of equations
is obtained iteratively.
This permits the calculation of the predicted or synthetic spectrum,
which is then compared to the observed UV spectrum.
The process is repeated with different stellar parameters
until a satisfactory fit is obtained.

The UV spectrum between 1000 and 2000\,\AA\ carries a lot of information:
P-Cygni-type profiles of resonance lines of several ions of
C, N, O, Si, S, P, as well as hundreds of strongly wind-contaminated
lines of Fe\,{\sc iv}, Fe\,{\sc v}, Fe\,{\sc vi}, Cr\,{\sc v},
Ni\,{\sc iv}, Ar\,{\sc v}, Ar\,{\sc vi}.
But the information about the stellar parameters can be extracted
only after careful analysis.
A very important recent improvement of our method concerns the
development of a substantially consistent treatment of the
blocking and blanketing influence of all metal lines in the entire
sub- and supersonically expanding atmosphere.
All the results we will present are based on this new
generation of models.

The operational procedure is as follows.
A preliminary inspection of the visual or UV spectrum of the star
to be analyzed gives an initial estimate of $\Teff$.
From the UV spectrum, the terminal wind velocity $\vinf$ can be
measured directly.
Now an initial value for the stellar radius $R$,
defined at a Rosseland optical depth of $2/3$, is assumed.
Using this $R$ and $\vinf$ we obtain an estimate of the mass
($\vinf$ scales with $\sqrt{M/R}$ as explained by the theory
of radiation driven winds).
With the current values of $R$, $\Teff$, $M$, and assuming a
set of abundances, we can solve the model atmosphere and
calculate the velocity field, the mass loss rate $\Mdot$,
and the synthetic spectrum.
Now this predicted spectrum is compared to the observed one.
If the fit is not satisfactory, we need to modify $\Mdot$ via
a change of $R$ (since $\log \Mdot \sim \log L$,
according to the radiation driven wind theory).
The change in $R$ forces us to change the mass, too,
in order to keep $\vinf$ consistent with the observed value.
The new model is calculated and the process is repeated until
we obtain a good fit to all features in the observed spectrum.

\begin{figure}[t]
\includegraphics[height=3.0in,angle=-90]{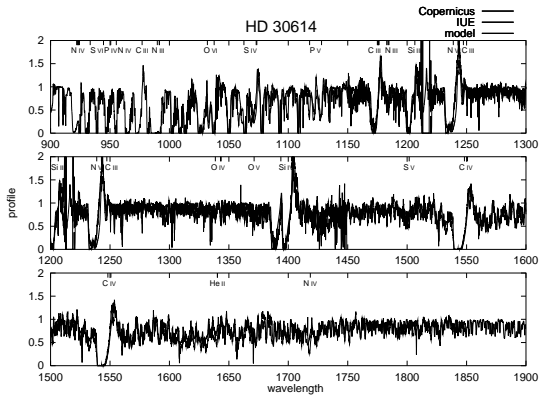}
\hfill\parbox[t]{2.5in}{
\caption{
Synthetic UV spectrum of a model for $\alpha$~Cam compared to
the observations by Copernicus and IUE,
demonstrating the quality that can be achieved with
our new model generation.
\label{fig:alphaCam}}}
\end{figure}

Figure~\ref{fig:alphaCam} shows the result of applying this
procedure to the O~supergiant HD~30614 ($\alpha$~Cam).
The final parameters are $\Teff=29000\,{\rm K}$, $\log g=3.0$,
$R=30\,\Rsun$, $\Mdot=4\times 10^{-6} \Msun\,{\rm yr}^{-1}$,
$\vinf=1500\,{\rm km/s}$.
(Note that the consistent treatment of the hydrodynamics in our
procedure will be described in a forthcoming paper
(Pauldrach and Hoffmann~2002); the method is also illustrated
in these proceedings by Hoffmann and Pauldrach.)
This implies a stellar mass of $33\,\Msun$ and a
spectroscopic distance of 1.2 kpc, in good agreement with the
distance estimate of 1 kpc (Scuderi et al.~1998).
Thus we show that our current models produce
satisfactory results for massive Population~I stars.
Can we apply the same procedure to CSPNs?

\section{The relation between wind-momentum loss rate and luminosity}

\begin{figure}[t]
\includegraphics[height=3.0in,angle=-90]{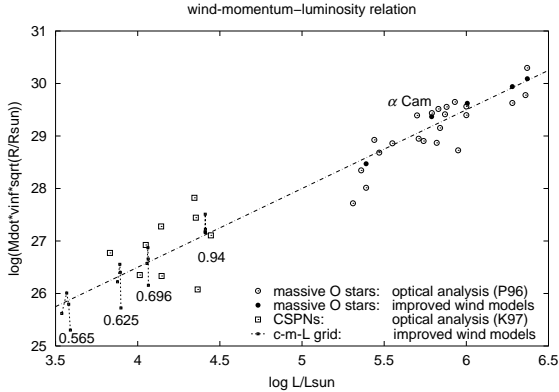}
\hspace{-0.1in}\hfill\parbox[t]{2.6in}{
\caption{
The wind-mo\-mentum--luminosity relation for massive O~stars
and CSPNs.
P96 designates the analysis based on
H$\alpha$ profiles by Puls et al.~1996,
K97 that of CSPNs by Kudritzki et al.~1997.
Also plotted are the calculated wind momenta for a
sample of massive O~stars and for
a grid of stars following post-AGB evolutionary tracks
(masses given in $\Msun$).
\label{fig:wml1}}}
\vspace{-3mm}
\end{figure}

The radiatively driven wind theory predicts, for fixed abundances,
a simple relation between the quantity $\Mdot \vinf$, which has
the dimensions of a momentum loss rate, and the stellar luminosity:
$$
\Mdot \vinf \sim R^{-0.5} L^{(1/\alpha)}
$$
where $\alpha$, the power law exponent of the line strength
distribution function, is \mbox{$\simeq 2/3$}
(slightly dependent on temperature and metallicity;
see, for example, Puls et al.~1996).
It is practical to plot the log of $\Mdot \vinf R^{0.5}$
as a function of $\log L$.
In this kind of plot the theory predicts, in first approximation,
a linear relation, which is indeed followed by all kinds of
massive hot stars, as shown in Figure~\ref{fig:wml1}.

An initial attempt to verify if CSPNs follow the
wind-momentum--luminosity relation was partly successful
(see Figure~3 in Kudritzki et al.~1997 and also our
Figure~\ref{fig:wml1}).
In that paper $Q$-values (a quantity relating mass loss rate
and stellar radius, $Q \sim \Mdot (R \vinf)^{-3/2}$)
were derived from observed H$\alpha$ profiles,
and the stellar masses were derived from $\Teff$ and $\log g$,
using post-AGB tracks plotted in the $\log g$--$\log \Teff$ diagram.
The stellar radii (and thus, mass loss rates) and luminosities
were then obtained from the masses and the post-AGB
mass--luminosity relation.
The CSPNs were found to be at the expected position along the
wind-momentum--luminosity relation, indicating a qualitatively
successful prediction by the theory of radiatively driven winds.
However, the situation was not satisfactory because
there appeared to be a large dispersion in wind strengths
at a given luminosity (strong-winded and weak-winded CSPNs) and
some of the CSPN masses and luminosities were very high
($M>0.8\,\Msun$),
in contradiction with theoretical post-AGB evolutionary speeds

Thus, at that point we had a qualitative positive result,
namely that in principle the CSPN winds obey the same
physics as the massive O star winds;
but we also had some unsolved problems which we now want to
rediscuss using the improved model atmospheres.

As a first step we have used our models to calculate the
terminal velocities and mass loss rates for a grid of stars
following the current theoretical post-AGB evolutionary tracks
(see, for instance, Bl\"ocker 1995);
the resulting wind momenta are also plotted in Figure~\ref{fig:wml1}.
The numerical models do nicely follow the expected
theoretical wind-momentum--luminosity relation,
although the spread is much smaller than that in the
values derived by Kudritzki et~al.
Furthermore, the location of the observed sample in the diagram
indicates masses between $0.6$ and $0.95\,\Msun$, with a clear
absence of CSPNs with masses below $0.6\,\Msun$.

\begin{figure}[t]
\includegraphics[height=3.0in,angle=-90]{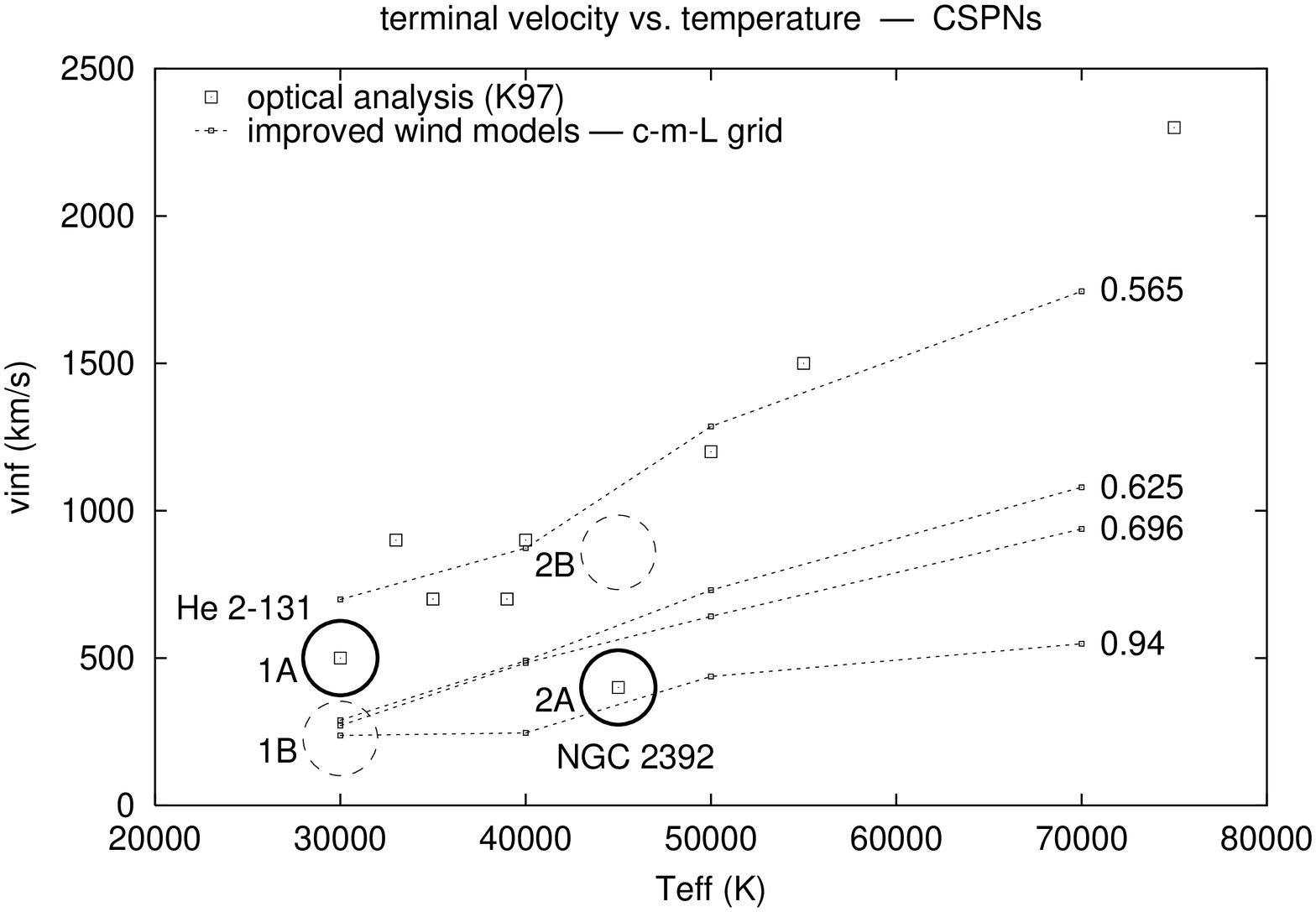}
\hfill\parbox[t]{2.5in}{
\caption{
Terminal velocities calculated for a grid of stars following
post-AGB evolutionary tracks (dashed lines, masses in $\Msun$
labelled on the right) compared to observed values (squares).
\label{fig:vinf}}}
\includegraphics[height=3.0in,angle=-90]{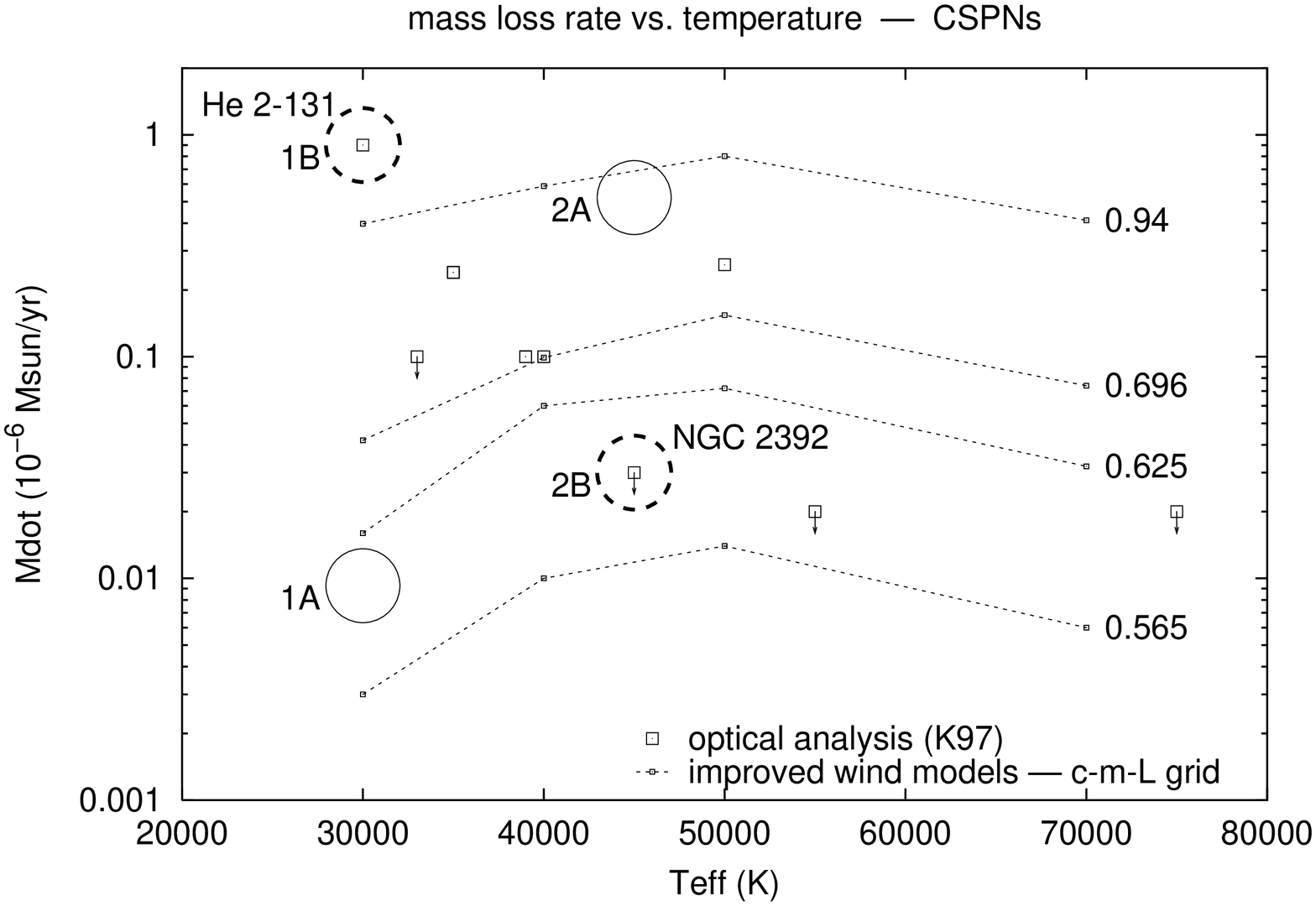}
\hfill\parbox[t]{2.5in}{
\caption{
Mass loss rates computed for the above grid of stars,
compared to the values derived by Kudritzki et al.\ (1997)
for the same set of observations as in Figure~\ref{fig:vinf}.
\label{fig:Mdot}}}
\end{figure}

This result is entirely unexpected from the
standpoint of current evolutionary theory.
To understand this discrepancy,
we must compare the relations of the individual
dynamic quantities, $\vinf$ and $\Mdot$.
This is done in Figures~\ref{fig:vinf} and~\ref{fig:Mdot}:
Figure~\ref{fig:vinf} shows our calculated terminal velocities
and the observed values,
Figure~\ref{fig:Mdot} shows our computed mass loss rates
and those derived by Kudritzki et~al.\ for their sample.

Here, too, a fundamental discrepancy immediately becomes obvious:
where\-as the positions of the observations in the
diagram showing the terminal velocities cluster at
rather small CSPN masses (between $0.5$ and
$0.6\,\Msun$), their mass loss rates point to a majority of
masses above $0.7\,\Msun$.

A detailed look at the positions of individual CSPNs in
the plots reveals even more alarming discrepancies.
Take, for example, He~2-131.
Its terminal velocity would indicate a mass of about
$0.6\,\Msun$ (circle 1A in Figure~\ref{fig:vinf}).
But this mass is completely irreconcilable with its mass loss
rate: it is found not at the position labelled 1A in
Figure~\ref{fig:Mdot}, but at 1B, with $\Mdot$ a factor of
hundred higher, suggesting a mass of above $0.94\,\Msun$!
The reverse is true for NGC~2392.
Its terminal velocity points to a mass of about $0.9\,\Msun$
(circle 2A in Figure~\ref{fig:vinf}), but its observed
mass loss rate is much too small for this mass (circle 2B
in Figure~\ref{fig:Mdot}), indicating a mass of approximately
$0.6\,\Msun$.

We still face a problem if we take the mass loss rate
determinations of Kudritzki et~al.\ to be correct:
then our calculations would place these two stars at the
positions labelled 1B and 2B in Figure~\ref{fig:vinf}
-- with terminal velocities differing by a factor of 2 to 3.
But this is clearly ruled out by the observations.
($\vinf$ is a directly measurable quantity.)

We are therefore left with the conclusion that the analysis of
Kudritzki et~al.\ is at odds with the mass loss rates computed
by our models based on the post-AGB evolutionary tracks.
To determine whether the reason for this discrepancy lies with the
evolutionary tracks on the one hand or our hydrodynamical models
and the analysis by Kudritzki et~al.\ on the other
requires further observational evidence.
This is given to us by the UV spectra of the stars.

\section{Examples: UV analysis of the CSPNs He~2-131 and NGC~2392}

\begin{figure}[p]
\includegraphics[height=0.49\textwidth,angle=90]{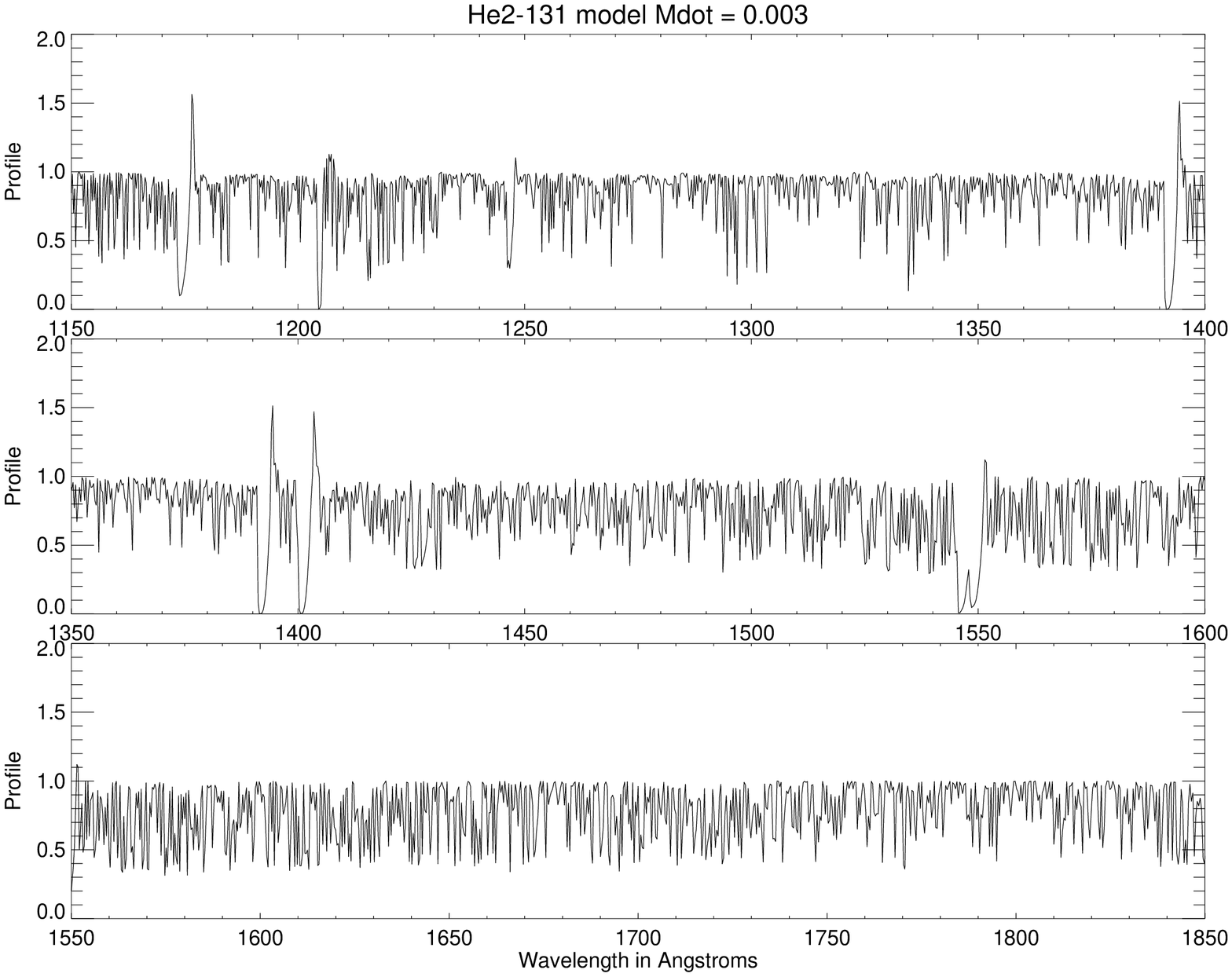}\hfil
\includegraphics[height=0.49\textwidth,angle=90]{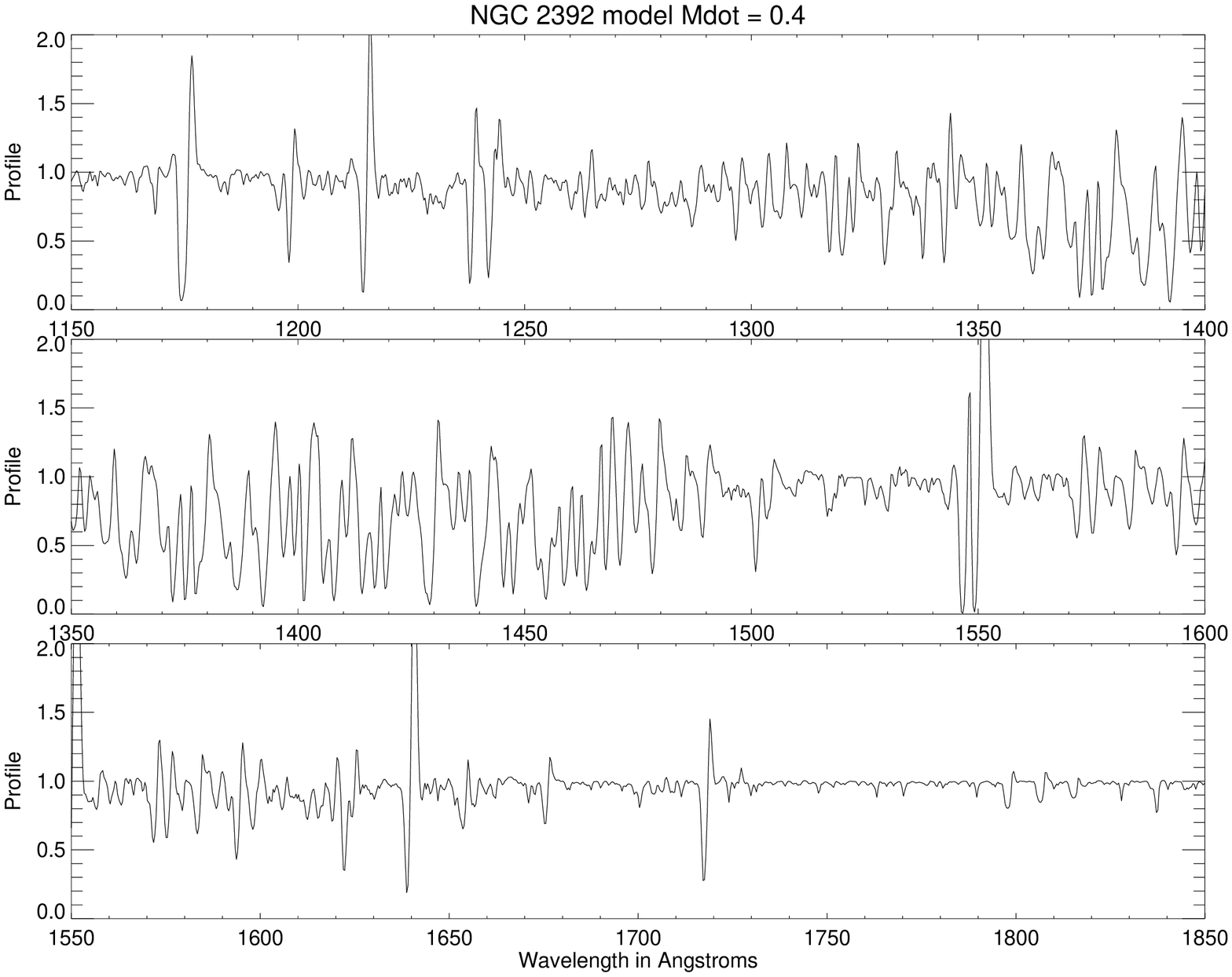}\\[5mm]
\includegraphics[height=0.49\textwidth,angle=90]{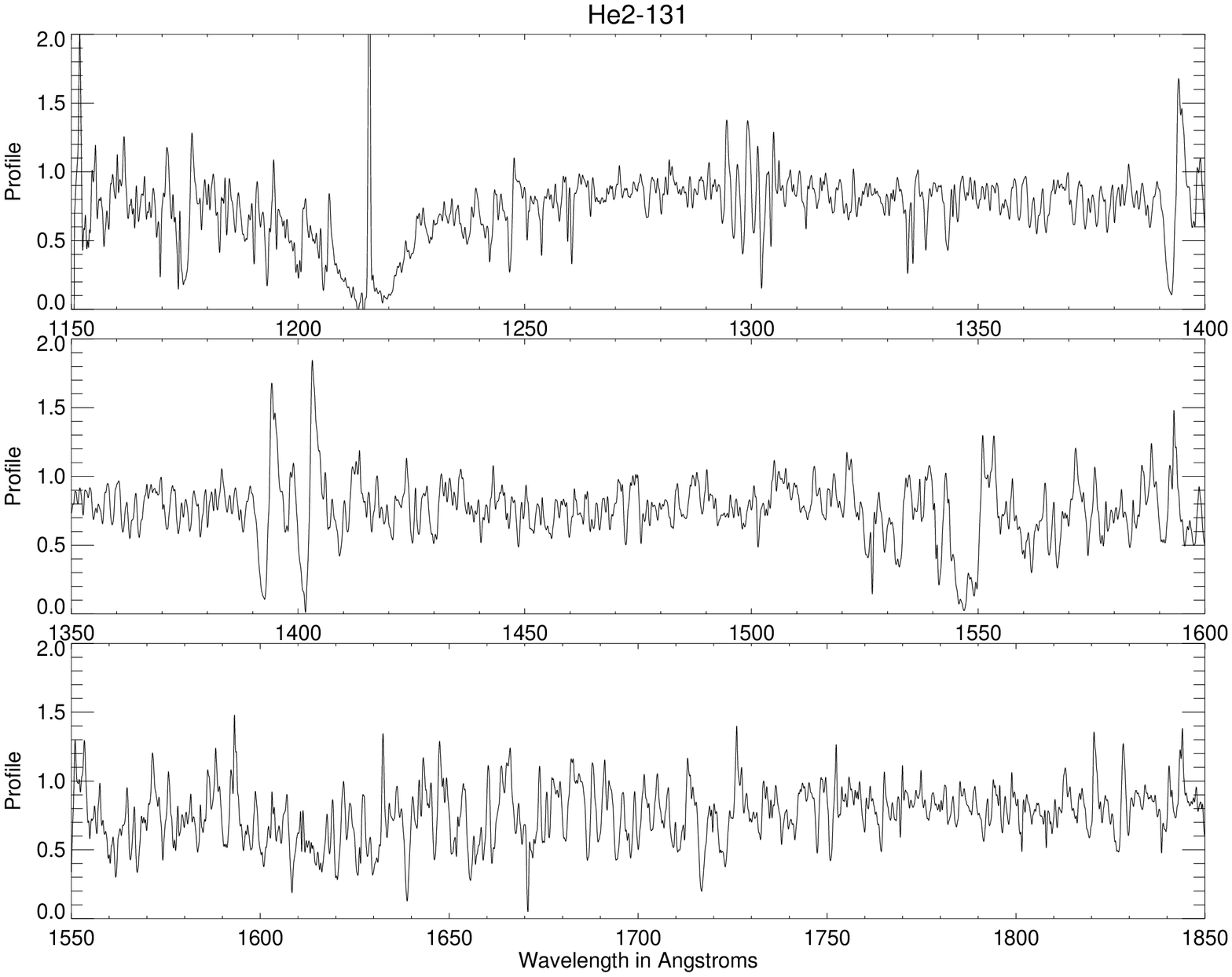}\hfil
\includegraphics[height=0.49\textwidth,angle=90]{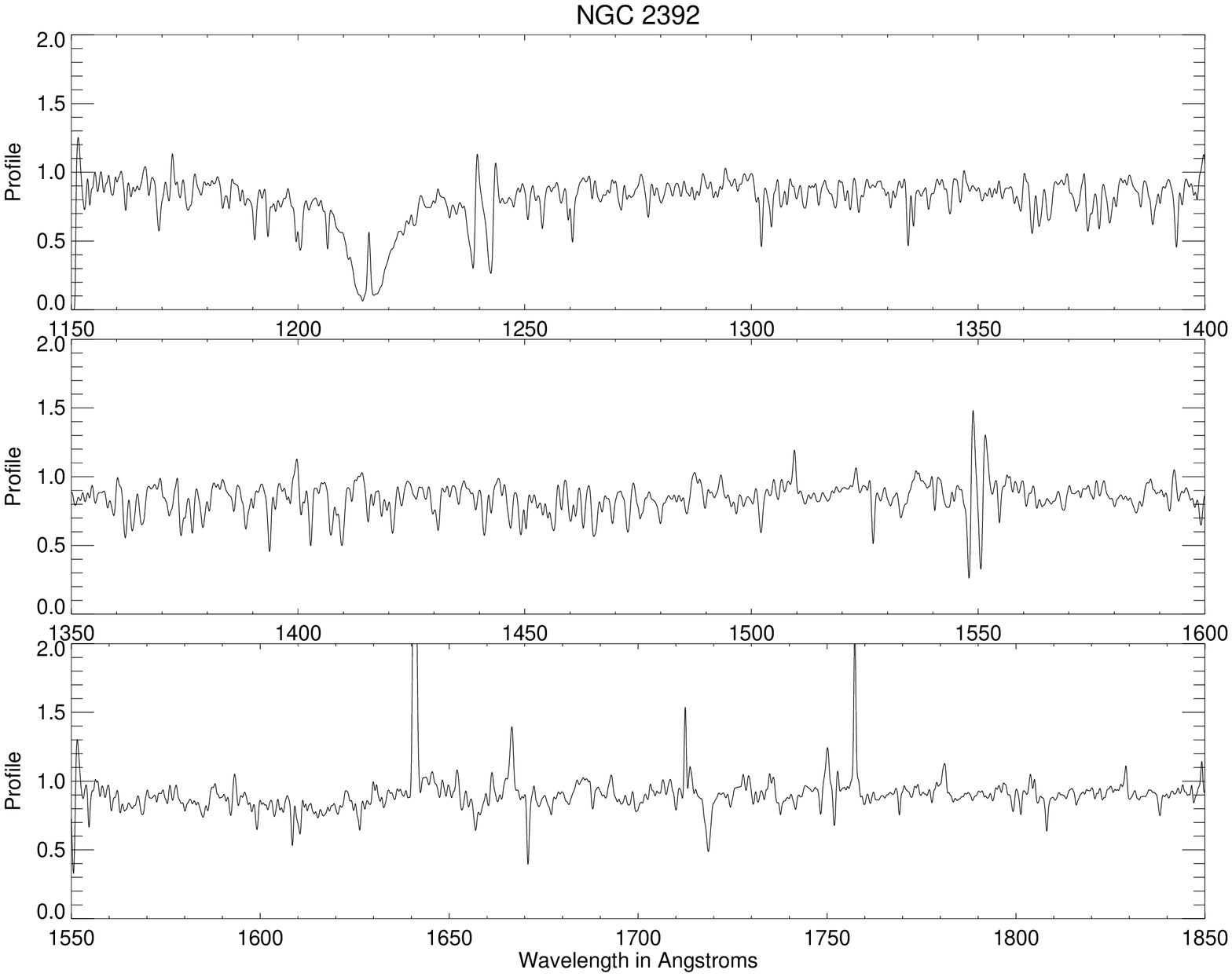}\\[5mm]
\includegraphics[height=0.49\textwidth,angle=90]{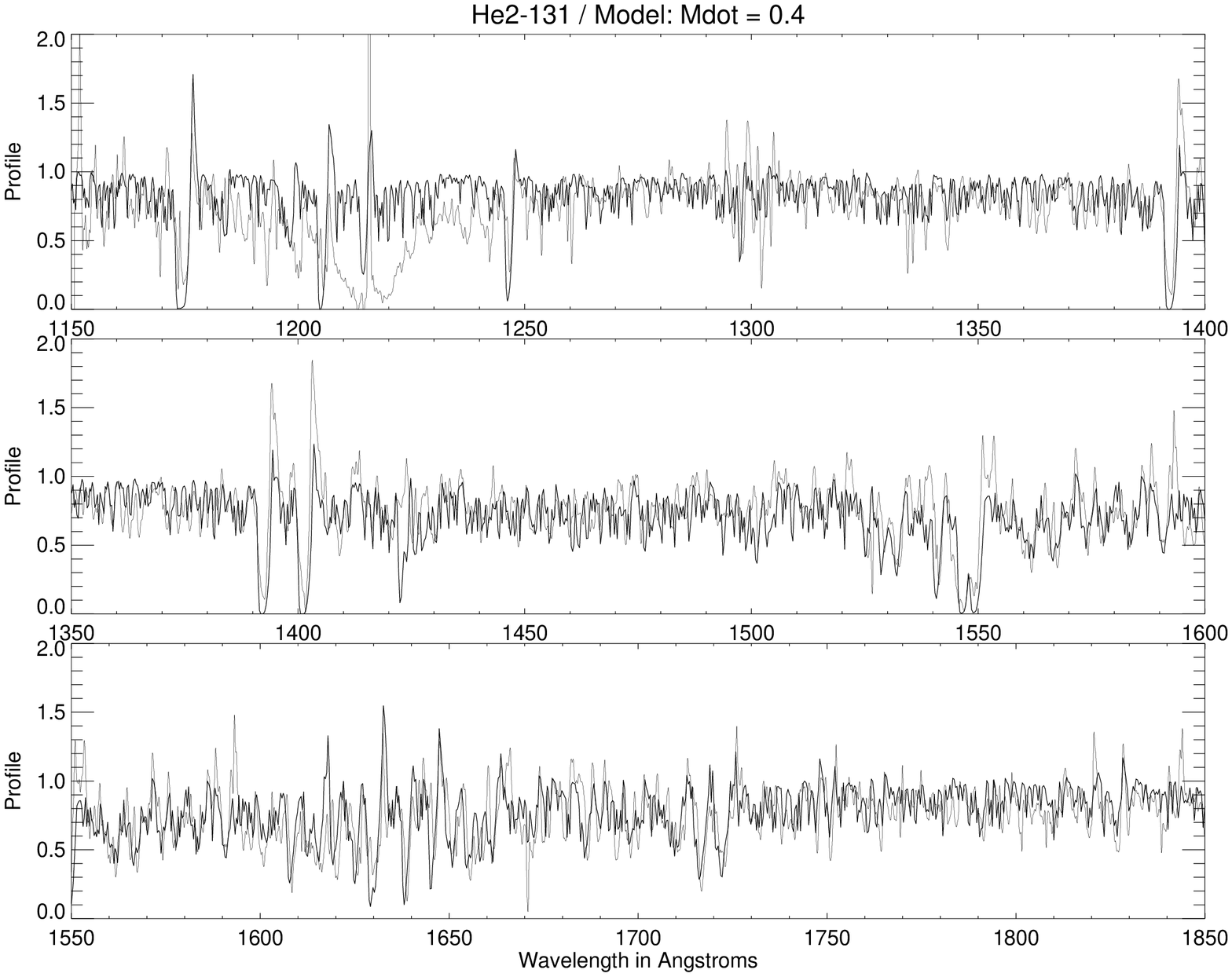}\hfil
\includegraphics[height=0.49\textwidth,angle=90]{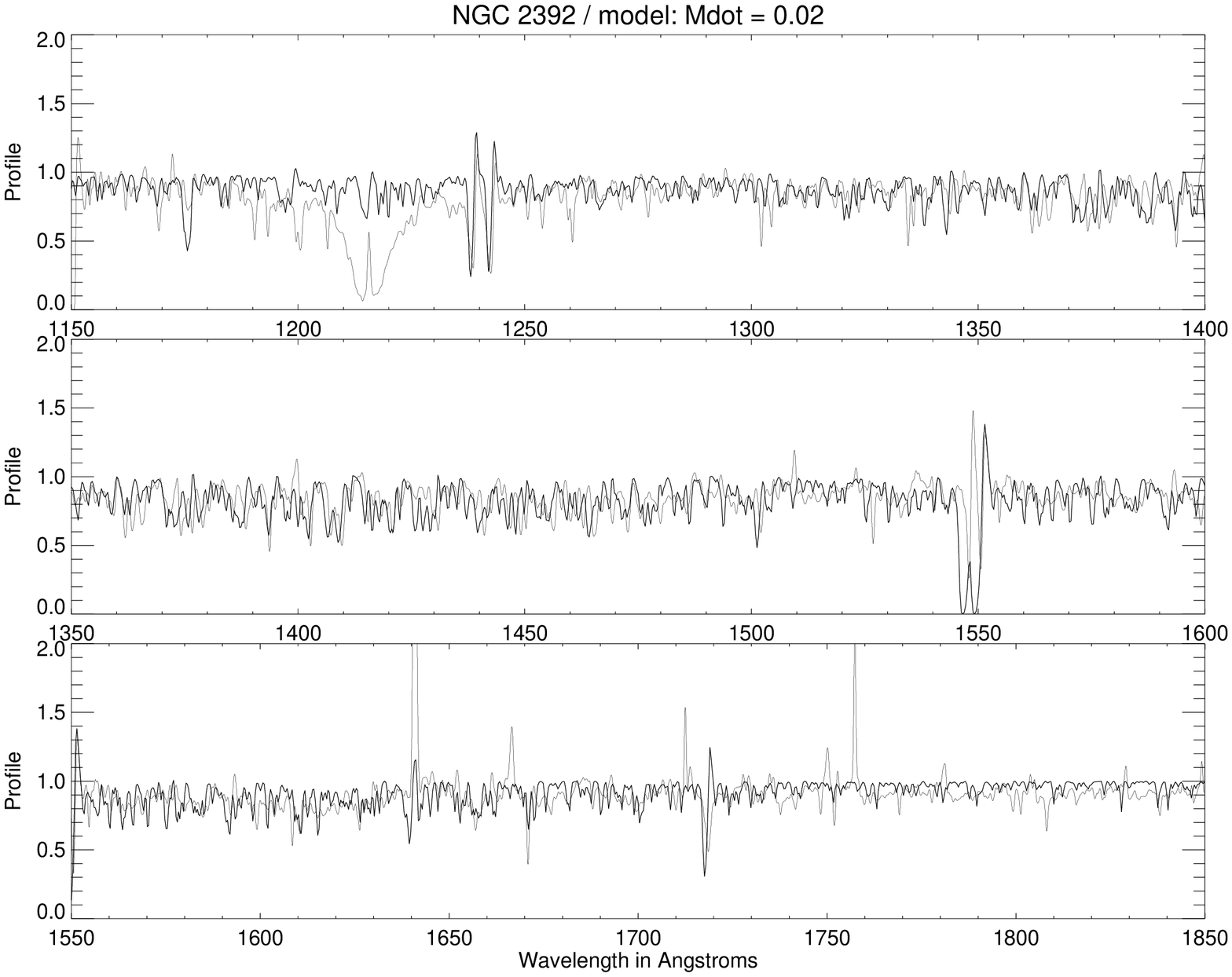}
\caption{
{\em (Left)\/}~Top:~Synthetic spectrum of model 1A for
\mbox{He~2-131} (see text).
This is incompatible with the observed IUE spectrum (middle).
A model with a significantly enhanced luminosity and thus
mass loss rate reproduces the distinctive features
in the UV spectrum much better (bottom, overplotted with the
observed spectrum to better show the similarity).
{\em (Right)\/}~Top:~Synthetic spectrum of model 2A for NGC~2392.
Again, this is incompatible with the observed IUE spectrum (middle).
In this case, however, the mass loss rate (and thus $L$) is much too high;
a model with lower luminosity reproduces the spectrum much
better (bottom).
\label{fig:spectra}}
\end{figure}

Figure~\ref{fig:spectra} (top left) shows the synthetic
UV spectrum of the model corresponding to 1A in
Figures~\ref{fig:vinf} and~\ref{fig:Mdot}.
It is clearly incompatible with the observed spectrum
of He~2-131 (middle), since its mass loss rate is obviously too
small, as evidenced by the presence of almost only purely
photospheric lines hardly influenced by the thin wind --
indicating that this CSPN must have a much larger luminosity,
because $L$ is the major factor determining the mass loss rate.
Thus, we have calculated a series of models with increasing
luminosity -- and therefore increasing mass loss rate --
(at the same time adjusting the mass to keep
the terminal velocity at its observed value)
to see whether one of these models could reproduce the
numerous strongly wind-contaminated iron lines observed especially
in between $1500$ and $1700\,{\rm\AA}$.
Indeed, a model which due to its luminosity yields
approximately the high mass loss rate of~1B
gives a much improved fit -- see Figure~\ref{fig:spectra},
bottom left.
(The parameters of this star are given in Table~\ref{tbl:params}.)

The situation is reversed with NGC~2392.
The synthetic spectrum of model 2A is incompatible with the
observed UV spectrum (Figure~\ref{fig:spectra}, right top and
middle),
since it produces many strongly wind-contaminated lines,
which are, however, not observed.
Instead, almost only photospheric lines are produced by the
star.
Again the problem is the luminosity, which in this case is much
too high.
Decreasing the luminosity and thus the mass loss rate yields a
model with a much better agreement with the observed spectrum
(Figure~\ref{fig:spectra}, bottom right).

What does this mean for the derived stellar parameters,
which by virtue of this analysis now have the status of
{\em observed quantities\/}?
Let us consider first the weak-winded CSPN, NGC~2392.
We determine a $\Teff$ of 40000\,K from the ionization
equilibrium of Fe ions in the stellar UV spectrum,
not too different from the value obtained by the
ionization equilibrium of He\,{\sc i} and He\,{\sc ii}
(absorption lines in the optical stellar spectrum).
The very low terminal velocity in the wind of
$400\,{\rm km\,s}^{-1}$ leads, together with a decreased
luminosity (in order to reduce the predicted mass loss rate
until the predicted and observed spectra are in good agreement),
to a small radius.
From this radius ($1.5\,\Rsun$) and $\vinf$
we get a stellar mass of only $0.41\,\Msun$,
a value much smaller than if we believe in the classical
mass--luminosity relation -- a high mass of $0.9\,\Msun$
was the result found by Kudritzki et~al.
(But due to the smaller $R$ and $\Teff$ our luminosity is also
smaller.)
We remark at this point that our error in the mass is
extremely small ($\le 0.1\,\Msun$) due to the sensitive dependence
on $\vinf$ and the small error in this value ($\le 10\%$).
Furthermore, our predicted values of $\vinf$ are in agreement
within 10\% with the observed values for the case of massive
O~stars (cf.~Hoffmann and Pauldrach, these proceedings).
Thus the systematic error is almost negligible.

Next we consider the central star of He 2-131.
In this case the terminal velocity of $500\,{\rm km\,s}^{-1}$
($\Teff=33000\,{\rm K}$) would appear to suggest,
according to the classical post-AGB mass--luminosity relation,
a stellar mass of about $0.6\,\Msun$ (cf.~Figure~\ref{fig:vinf}).
However, the wind features observed in the UV spectrum
forced us to increase the stellar $R$ and $L$,
which in turn increased $\Mdot$ until a good fit was obtained.
From the corresponding large radius -- $5.5\,\Rsun$ -- and
$\vinf$ we derive a stellar mass of $1.39\,\Msun$,
a value very close to the Chandrasekhar mass
of relevance for type~Ia supernovae.
Thus, in this case the resulting mass is even more extreme than
the value of $0.9\,\Msun$ obtained by Kudritzki et~al.

\section{Interpretation of CSPN winds: results and discussion}

\begin{table}[t]
\caption{Parameters of eight CSPNs derived by our analysis of
the UV spectra using our model atmospheres,
compared to the values found by Kudritzki et al.~1997.
\label{tbl:params}}
\begin{center} \small
\begin{tabular}{lccccr@{.}l@{\extracolsep{5pt}}r}
& $\Teff$ & $R$ & & $M$ &
\multicolumn{2}{c}{$\Mdot$} &
\multicolumn{1}{c}{$\vinf$} \\
\multicolumn{1}{c}{\raisebox{7pt}[-7pt]{Object}} & (K) & ($\Rsun$) &
\raisebox{7pt}[-7pt]{$\displaystyle \log {L \over \Lsun}$} & ($\Msun$) &
\multicolumn{2}{@{}c@{}}{($10^{-6}\Msun/{\rm yr}$)} &
\multicolumn{1}{@{}c@{}}{(km/s)} \\[2pt] \hline
\multicolumn{8}{c}{our models} \\ \hline
NGC 2392 & 40000 & 1.5 & 3.7 & 0.41 &       0&018 &  420 \\
NGC 3242 & 75000 & 0.3 & 3.5 & 0.53 &       0&004 & 2400 \\
IC 4637  & 55000 & 0.8 & 3.7 & 0.87 &       0&019 & 1500 \\
IC 4593  & 40000 & 2.2 & 4.0 & 1.11 &       0&062 &  850 \\
He 2-108 & 39000 & 2.7 & 4.2 & 1.33 &       0&072 &  800 \\
Tc 1     & 35000 & 3.0 & 4.1 & 1.37 &       0&021 &  900 \\
He 2-131 & 33000 & 5.5 & 4.5 & 1.39 &       0&35  &  450 \\
NGC 6826 & 44000 & 2.2 & 4.2 & 1.40 &       0&18  & 1200 \\ \hline
\multicolumn{8}{c}{Kudritzki et al.~1997} \\ \hline
NGC 2392 & 45000 & 2.5 & 4.4 & 0.91 & ~$\le$ 0&03  &  400 \\
NGC 3242 & 75000 & 0.6 & 4.0 & 0.66 & $\le$ 0&02  & 2300 \\
IC 4637  & 55000 & 1.3 & 4.1 & 0.78 & $\le$ 0&02  & 1500 \\
IC 4593  & 40000 & 2.2 & 4.0 & 0.70 &       0&1   &  900 \\
He 2-108 & 35000 & 3.2 & 4.1 & 0.75 &       0&24  &  700 \\
Tc 1     & 33000 & 5.1 & 4.4 & 0.95 & $\le$ 0&1   &  900 \\
He 2-131 & 30000 & 5.5 & 4.3 & 0.88 &       0&9   &  500 \\
NGC 6826 & 50000 & 2.0 & 4.3 & 0.92 &       0&26  & 1200 \\ \hline
\end{tabular}
 \end{center}
\end{table}

Table~\ref{tbl:params} shows the result of applying the
method and UV analysis described in the previous section
to eight CSPNs, including the two exemplary objects above.
The problem which Table~\ref{tbl:params} points to is obvious:
according to the post-AGB evolutionary timescales we should not
find so many extremely luminous CSPNs, because according to this
theory they are expected to fade very quickly.
We shall note, however, that the sample of objects
chosen here is most likely not a representative one.

\begin{figure}[t]
\centerline{\includegraphics[height=4.1in,angle=-90]{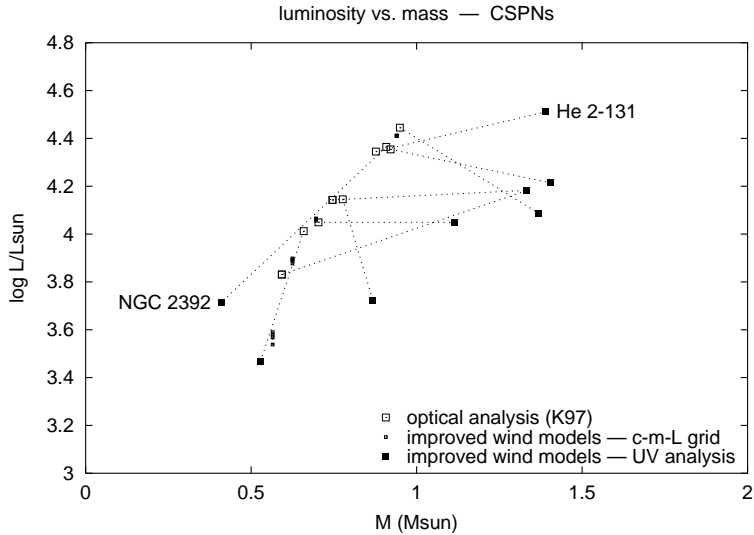}}
\caption{
Luminosity vs.\ mass for the evolutionary tracks (open squares)
compared to the observed quantities determined with our method
(filled squares).
Although the luminosities deduced from the UV spectra lie in
the expected range, a much larger spread in the masses
(from $0.4$ to $1.4\,\Msun$) is obtained.
No well-defined relation between CSPN mass and luminosity
can be made out.
\label{fig:ml}}
\end{figure}

Figure~\ref{fig:ml} shows the relation between stellar mass
and luminosity obtained from our model atmosphere analyses,
in comparison with the mass--luminosity relation of the
evolutionary tracks, represented by the values from Kudritzki et
al.~1997.
This plot shows that the problem already indicated
is indeed very disturbing:
the derived masses and luminosities do not agree with the
classical post-AGB mass--luminosity relation.
{\em There is a very large spread in masses,
between 0.4 and 1.4\,$\Msun$,
and there is no well-defined relation
between CSPN mass and luminosity.}

\begin{figure}[t]
\includegraphics[height=3.0in,angle=-90]{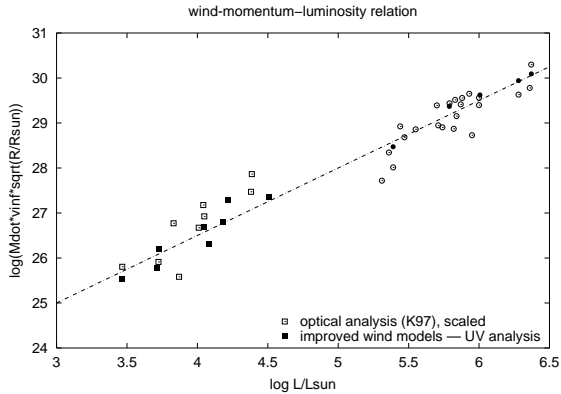}
\hfill\parbox[t]{2.5in}{
\caption{
The wind-mo\-mentum--luminosity relation for CSPNs (lower left)
based on our values determined from the UV spectra (filled squares).
The open squares are the values from Kudritzki et~al.\ with our
radii applied (see text).
Compared to Figure~\ref{fig:wml1} the result is striking.
\label{fig:wml2}}}
\end{figure}

In Figure~\ref{fig:wml2} we now again show the
wind-momentum--luminosity relation for both
massive hot stars and CSPNs, but this time based on the
parameters derived in our analysis.
Our models give wind momenta of the right order of
magnitude and within the expected luminosity range
(there may be too many CSPNs at $\log L/\Lsun > 4$,
but not so many as in Kudritzki et al.~1997).
The CSPNs are found along the extrapolation of the
wind-momentum--luminosity relation defined by the massive hot stars,
and the CSPNs show a smaller dispersion, i.\,e.,
a tighter correlation of wind-momentum with luminosity,
than was the case in Kudritzki et al.~(1997).
And, most important, this was achieved by fitting the multitude
of diagnostic features in the CSPN UV spectra by means of
up-to-date hydrodynamically consistent models.

How then does our wind-momentum--luminosity relation
compare to that found by Kudritzki et~al.?
The answer is, quite favorably.
If we drop the assumption made by Kudritzki et~al.\ that the
stars obey the theoretical post-AGB mass--luminosity relation,
and instead scale their mass loss rates to our radii\footnote{
Additionally allowing for their different effective temperatures
by requiring that the observed visual flux ($\sim R^2 \Teff$)
stay constant.}
-- keeping $Q$, the real observational quantity, fixed --
then their wind momenta match ours to within about
a factor of two.
Furthermore, their sample with the radii thus scaled
now also shows a much tighter correlation
of the wind momentum to luminosity than before
(see Figure~\ref{fig:wml2}).

All this is strong evidence that all these winds are radiatively driven.
It would be extremely difficult to explain the
wind-momentum--luminosity relation if there was
another mechanism driving the winds.

What makes the {\em CSPN mass discrepancy problem\/} found
most intriguing is the fact that the same model atmospheres,
with exactly the same physics, obviously work also perfectly well
for massive O stars, as we have shown.
We are reluctant to conclude that there is something basic
we do not understand about either the winds and photospheres
of O-type stars, or how to produce CSPNs and what their
internal structure is.
Both alternatives are difficult to believe.
But we need to explain why we obtain such a spread in the masses.

We cannot offer a fair solution to this paradox;
all we can do right now is to present our method, the results
obtained from our analysis, and the corresponding problem
in the clearest possible way, which is usually the
first step along the road that leads to the solution.

\end{document}